\documentclass[10pt]{article}
\usepackage[latin1]{inputenc}
\usepackage{a4wide}
\usepackage{graphicx}
\usepackage{amsmath}
\usepackage{mathrsfs}
\usepackage{amssymb}
\usepackage{amsfonts}

\begin{document}\title{An incomplete Kochen-Specker colouring}
\author{\small Helena Granstr\" om,\\ \small Department of Mathematics/Department of Physics, Stockholm University\\}

\maketitle 

\begin{abstract}

\noindent 
A particular incomplete Kochen-Specker colouring, suggested by Appleby in dimension three, is generalized to arbitrary dimension. We investigate its effectivity as a function of dimension, using two different measures. A limit is derived for the fraction of the sphere that can be coloured using the generalized Appleby construction as the number of dimensions approaches infinity. The second, and physically more relevant measure of effectivity, is to look at the fraction of properly coloured ON-bases. Using this measure, we derive a 'lower bound for the upper bound' in three and four real dimensions.

\end{abstract}

\newpage

\section{Introduction}
The Kochen-Specker theorem \cite{KS} is a result that has proved to be of great conceptual interest to quantum mechanics and its interpretation.
Let $f$ be a function from the set of projection operators in some Hilbert space to the set $\lbrace 0,1 \rbrace$ such that \begin{equation} \label{proj} \sum_{i=1}^N f(P_i)=1 \end{equation} where the $P_i$ are projection operators associated to the vectors $\mid e_i \rangle$, forming an ON-basis for the Hilbert space, and $N$ is the dimension of the space.\\
The statement of the Kochen-Specker theorem is that no such truth value assignments exist if the dimension of the Hilbert space is larger than two. This result is often translated in terms of two-colourings of spheres, and the same terminology will be used here.
The statement of KS (in $N$ real dimensions) is in these terms that no complete colouring of the $N-1$-sphere obeying the KS criteria (\ref{proj}) is possible. The question then arises how close to complete we can possibly get.
A collection of rather eccentric colourings 'almost' satisfying the KS criteria have been proposed by Pitowsky \cite{P}, Meyer \cite{M}, Kent \cite{K} and Clifton and Kent \cite{CK}.\\
Partly in response to these constructions and the interpretations claimed for them, Appleby~\cite{Appleby} in a 2003 paper derives an upper bound for the effectivity of a colouring that, in contrast to the colourings mentioned above, is measurable and hence arguably of greater physical interest.\\   
Appleby considers the question of maximal effectivity in the case of three-dimensional real Hilbert space, and argues that the maximal fraction of $S^{2}$ that can be satisfactorily KS coloured might be as large as $99 \%$. To provide a lower bound for the maximal effectivity in terms of area of $S^{2}$ coloured, Appleby also suggests a specific incomplete colouring that covers $87 \%$ of the sphere.\\
Here, the Appleby construction will be generalized to an arbitrary number of real dimensions. For $N=4$ the colouring analogous to that proposed by Appleby in dimension three covers  $79 \% $ of $S^{N-1}$, $74 \%$ for $N=5$ and $71 \%$ for $N=6$. The integer giving the least percentage is $N=12$; about $66.76 \%$.\\
A natural question to ask is what happens to this percentage as the number of dimensions gets very large. Does the colourable fraction, using this specific construction, go to zero? In fact it does not, but instead tends to $68 \%$ as $N$ approaches infinity.\\
A possibly more physically relevant way of phrasing the question of effectivity is in terms of the fraction of all possible ON-bases that can be coloured. In the following, we will look at the cases $N=3$ and $N=4$. The percentages turn out to be $69 \% $ and $34 \%$, respectively. The fact that the falloff is considerably larger than was the case for vectors suggests that this fraction might go to zero with an increase in the number of dimensions.\\
Several questions remain unanswered, however. Perhaps most importantly, our discussion is limited to families of pure states whose components are all real in some suitable basis in complex Hilbert space, and  different results might be obtained if this restriction were lifted. Secondly, one would want to evaluate the effectivity of the colouring in arbitrary dimension, using the fraction of colourable bases as a measure. One could also try to experiment with other constructions to further sharpen the lower bound for the upper bound provided by specific examples.

\section{An incomplete Kochen-Specker colouring} 
\begin{figure}[h]
\label{r01}
\begin{center}
\includegraphics [width=0.2\textwidth]{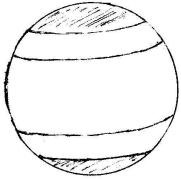}
\caption{A possible (incomplete) KS colouring of the unit two-sphere.} \label{KSsphere}
\end{center}
\end{figure}

Let us first consider the three-dimensional case and then go on from there to higher dimensions. We are interested in assigning the value $0$ or $1$ to vectors in $\mathscr{H}^3$ in such a way that no set of three mutually orthogonal vectors are all assigned the value $0$, and no pair of orthogonal vectors both have the value $1$. More precisely, we will colour the one-dimensional subspaces spanned by vectors. These conditions can be expressed as
\begin{equation} \label{defg} f: S^2 \to \lbrace 0,1 \rbrace \end{equation} 
\begin{equation}  f(P_1) + f(P_2) + f(P_3) = 1 \label{KSrule} \end{equation} 
for all sets of orthogonal vectors $ \lbrace P_1, P_2, P_3 \rbrace$, $S^2$ being the unit two-sphere. 
Letting white represent the value $0$ and black the value $1$, this problem can be translated into the problem of colouring $S^2$, in a way that satisfies the conditions just stated. In particular, antipodal points will have the same colour.\\
Any such assignment of truth values (probabilites from $\lbrace 0,1 \rbrace$) to all vectors in the Hilbert space of some system would correspond to (the possibility of) the system having well-defined properties, independent of measurement context. That is, for any possible observable the outcome of the corresponding measurement would be fully determined in advance. However, by the Kochen-Specker theorem, a complete such assigment of truth values is impossible. Hence, what we will do is to assign probabilites from $\lbrace 0,1 \rbrace$ according to (\ref{KSrule}) to some of the vectors in $\mathscr{H}$ - some vectors will necessarily remain uncoloured, by KS.\\
The way to go about this suggested by Appleby, is to start out by colouring the two polar caps defined by $\mid \tan \theta \mid < 1$ black, and the region around the equator bounded by $\mid \tan \theta \mid = \sqrt{2}$ white, where $\theta$ is the usual polar angle. This type of colouring is sketched in figure~\ref{KSsphere}.
\\
These limits are derived as follows. The two polar caps are made small enough so that no two vectors in an orthogonal triple can simultaneously lie in the black region, which means that they will extend down to $\theta = \frac{\pi}{4}$. The white section around the equator is just wide enough so that not all three vectors can lie in it at the same time.\\
Already in four dimensions the contribution to the total area by the black cap is close to negligible. As we will see below it reduces further with increasing dimension, which is why we will primarily be interested in the area taken up by the white section.\\ 
The fraction of the sphere in $N$ dimensions that can be coloured white with the given restriction is
\begin{equation} \label{white} F = \frac{\int^{\frac{\pi}{2}}_{ \arcsin{ \sqrt{ \frac{N-1}{N} } } }\sin^{N-2}
{\theta}d\theta }  {\int^{ \frac{\pi}{2} }_{0}\sin^{N-2}{\theta}d\theta}= 
2\frac{\textrm{vol}(S^{N-2})}{\textrm{vol}(S^{N-1})}\int^{\frac{\pi}{2}}_{\arcsin{\sqrt{\frac{N-1}{N}}}}\sin^{N-2}{\theta}d\theta \end{equation}
where vol$(S^d)$ denotes the surface area of the $d$-dimensional sphere. The intergral limits are derived using the expression  
\begin{equation}\label{radius} R_n= \sqrt{\frac{n}{n+1}}= \sqrt{\frac{N-1}{N}} \end{equation} for the radius of the circumsphere of a regular $n$-simplex, where $n=N-1$ is the dimension of the sphere in $N$ dimensions.\\   \\
As for the black area, $B_N$, it will in analogy with the $N=3$ case be located around the poles of the sphere, with limiting angle $\frac{\pi}{4}$;
\begin{equation} B_N = \textrm{vol}(S^{N-2})\int^{\frac{\pi}{4}}_{0}\sin^{N-2}{\theta}d\theta. \end{equation}  
What, one may ask, is the fraction of the sphere in $N$ dimensions that can be coloured using this method in the limit $N \to \infty$?
As can be seen from the expression
\begin{equation} \textrm{vol}(S^{d}) = \textrm{vol}(S^{d-1}) \int^{\pi}_0 \sin^{d-1} \theta d \theta \end{equation}
for high dimensions, the fraction of the area of the sphere that will lie around the poles is negligible, due to the increasingly sharp peak around $\theta = \frac{\pi}{2}$ of the sine function power. Thus the fraction of the surface area taken up by the black section will be very small. \\ 
To determine the fraction of the sphere taken up by the white section requires a bit more careful analysis. We will need to evaluate the expression
\begin{equation} \label{limit1} \lim_{N \to \infty} 2 \frac{ \textrm{vol}(S^{N-2}) } { \textrm{vol}(S^{N-1}) }
\int^{\frac{\pi}{2}}_{ \arcsin{ \sqrt{ \frac{N-1}{N} } } }\sin^{N-2}{\theta}d\theta. \end{equation} 
Using the known formula for $\textrm{vol}(S^d)$ we find that
\begin{equation} \frac{\textrm{vol}(S^{N-2})}{\textrm{vol}(S^{N-1})} = \frac{1}{\sqrt{\pi}} \frac{\Gamma(\frac{N}{2})}{\Gamma (\frac{N-1}{2})}. \end{equation}
This tends to $\sqrt{\frac{N}{2 \pi}}$ as $ N \to \infty $.\\
Next, let us take a look at the behaviour of the integral 
\begin{equation} \int^{\frac{\pi}{2}}_{\arcsin{\sqrt{\frac{N-1}{N}}}}\sin^{N-2}{\theta}d\theta = \int_{0}^{\arccos{\sqrt{\frac{N-1}{N}}}}\cos^{N-2}{\theta}d\theta \end{equation} when $N$ grows large. The second form is convenient because all expansions can be done around zero.\\
In the limit of large $N$ we can use
\begin{equation}\label{arccos} \arccos{ \sqrt{ \frac{N-1}{N} } } = \frac{1}{\sqrt{N}} + O( \frac{ 1 }{ N^{ \frac{3}{2} } } ). \end{equation}

\begin{figure}[h]
\begin{center}
\includegraphics [width=0.4\textwidth]{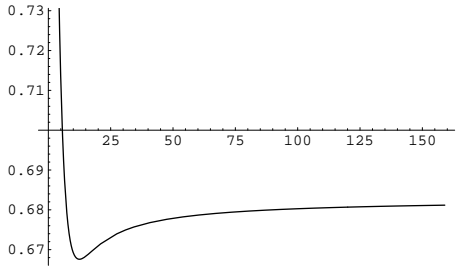}
\caption{Percentage of the sphere in $N$ dimensions that is colourable using the above method, as a function of $N$.} \label{plotN}
\end{center}
\end{figure}

Expanding $\cos t$ around $t=0$ and using the regular binomial expansion and the fact that when $N$ is large $N-2$ can be approximated with $N$,
\begin{equation} \lim_{N \to \infty}\cos^{N-2}{\theta} = (1-\frac{\theta^2}{2})^N + h(\theta, N)=  h(\theta, N) + 1 - N \frac{\theta^2}{2} + \frac{N^2}{2!}\frac{\theta^4}{4} - \frac{N^3}{3!} \frac{\theta^6}{8}+... \end{equation} 
where $h(\theta, N)$ is a function such that \begin{equation}  \lim_{N \to \infty} \sqrt{N} \int_{0}^{\arccos{\sqrt{\frac{N-1}{N}}}}h(\theta, N) = 0.  \end{equation} 
Integrating term by term and using the cosine expansion and equation~(\ref{arccos}) for the expansion of arccosine, we get
\begin{equation}  \label{goodsum} \lim_{N \to \infty} \int_{0}^{\arccos{\sqrt{\frac{N-1}{N}}}}\cos^{N-2}{\theta}d\theta  = \frac{1}{\sqrt{N}} \sum_{k=0}^{\infty} \frac {1}{2^k} \frac{1}{k!} \frac{(-1)^k}{(2k+1)} = \sqrt{\frac{\pi}{{2N}}}\textrm{erf}(\frac{1}{\sqrt{2}})\end{equation} with erf the statistic-probabilistic error function;
\begin{equation} \textrm{erf}(z) \equiv \frac{2}{\sqrt{\pi}}\int ^z_0 \textrm{e}^{-t^2} dt. \end{equation} \\
Putting all of this together, we have the result
\begin{equation} \lim_{N \to \infty} 2 \frac{ \textrm{vol}( S^{n-2} ) } { \textrm{vol}( S^{n-1} ) }
\int^{ \frac{ \pi }{ 2 } }_{ \arcsin{ \sqrt{ \frac{N-1}{N} } } } \sin^{N-2}{\theta}d\theta = 
\textrm{erf}( \frac{1}{ \sqrt{2} } ) \approx 0.68. \end{equation} \\
So, approaching the limit of an infinite number of dimensions of the Hilbert space $\mathscr H$ in which our projective measurements are conducted, binary probabilities (corresponding to well-defined, non-contextual properties of the system with available states in $\mathscr H$) can be assigned to approximately $68 \%$ of the vectors in $\mathscr H$.\\The behaviour of the percentage as a function of dimension is given in figure~\ref{plotN}. The results are $ 1 - \frac{1}{ \sqrt{2} }+ \frac{ 1 }{ \sqrt{3} } = 87\%$ of all vectors for $N=3$, $79 \% $ for $N=4$, $74 \%$ for $N=5$ and $71 \%$ for $N=6$. The integer giving the least percentage is $N=12$; about $66.76 \%$.
\\
What has been derived above is a lower bound for the area of the sphere that is KS colourable in arbitrary dimension. The possibility remains, however, that a maximally effective colouring could cover a much larger area - possibly, in fact, as much as $99 \%$ of the sphere in $\mathbb{R}^3$, see \cite{Appleby}. 

\section{A second effectivity measure}
\begin{figure}[h]
\begin{center}
\includegraphics [width=0.25\textwidth]{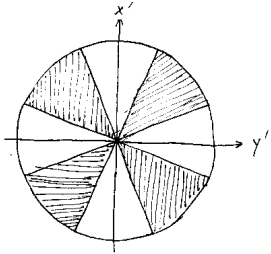}
\caption{The overlap between the white section of the great circle, and its rotation by $\frac{\pi}{2}$ is the shaded section. }  \label{cirkelsnitt2}
\end{center}
\end{figure}

The physically relevant question is, arguably, not how large a fraction of all states can be assigned probabilities $1$ or $0$, but rather what percentage of all complete orthogonal bases (measurements) can have all their basis vectors assigned binary probabilities in a consistent way. We will answer this question specifically for the Appleby colouring in three and four dimensions.\\
Let us first consider the colouring of the two-sphere proposed above - a black cap and a white equatorial belt covering in total $87\%$ of the sphere - and make use of the regular measure on $\mathbb{R}^3$ to compare the number of properly coloured bases consisting of vectors from these sections with the total number of ordered orthonormal triples in $\mathbb{R}^3$.\\
In a properly coloured base exactly one vector is black, so one of the three vectors in an orthogonal triple has to be chosen to lie on one of the black caps. The remaining two orthogonal vectors can then be chosen from a great circle orthogonal to the first vector - the question is how large a fraction of this great circle will lie within the white section and also how the second vector (which determines the third basis vector up to a sign) can be chosen so that the third vector will also be contained within the white section. \\
\begin{figure}[h]
\begin{center}
\includegraphics [width=0.2\textwidth]{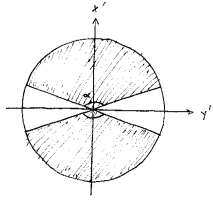}
\caption{A cut through the plane of the great circle orthogonal to the vector chosen to lie on the black cap. The intersection with the white area is shaded.} \label{cirkelsnitt1}
\end{center}
\end{figure}\\
Figure~\ref{cirkelsnitt1} depicts the plane of the great circle orthogonal to the first (black) vector on which the remaining two vectors in the orthogonal triple will have to lie. The circle segment bounding the shaded area is the cut between the white belt and the orthogonal great circle. For any choice of second vector from this section, the third vector will be fully determined (up to a sign). Hence, we cannot choose our second vector in a properly coloured triple from any part of the circle-belt overlap in figure~\ref{cirkelsnitt1}, but only from the sectors that will result in the third vector lying in the white belt as well. Given a second vector, the third is obtained by rotation in the great circle plane by an angle of $\frac{\pi}{2}$. The allowed choices for second vector are then the points such that the points corresponding to a $\frac{\pi}{2}$ rotation of these points are also white. This set of points is just the overlap between the white (shaded) sector in figure~\ref{cirkelsnitt1} and the same sector rotated by $\frac{\pi}{2}$, as illustrated in figure~\ref{cirkelsnitt2}, an overlap that can be shown to always be non-empty. Hence, what we will need to find is the total angle taken up by the shaded section in figure~\ref{cirkelsnitt2} - this will be denoted by $\beta$.
It is clear that this $\beta$ can be expressed in terms of the $\alpha$ of figure~\ref{cirkelsnitt1} as
\begin{equation} \beta = 4 \alpha - 2 \pi. \end{equation}
\\
 
\begin{figure}[h]
\begin{center}
\includegraphics [width=0.3\textwidth]{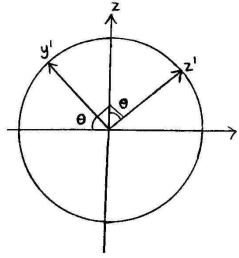} 
\caption{The vector $y'$ will make an angle $\pi - \theta$ with the $z$ axis.} \label{cutplane}
\end{center}
\end{figure}

\begin{figure}[h]
\begin{center}
\includegraphics [width=0.2\textwidth]{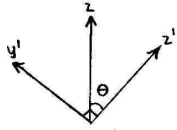} 
\caption{$y'$ lies on the two-sphere orthogonal to the vector $z'$.} \label{4D}
\end{center}
\end{figure}

The angle $\alpha$, in turn, can be expressed in terms of the regular polar angle $\theta$ that specifies our choice of black vector using the following procedure.\\ First, consider the plane spanned by the vector chosen to lie in the black section, call it $z'$, and a vector $y'$ in the plane orthogonal to $z'$; $\lbrace x, y, z \rbrace$ is a reference coordinate system as shown. The vector $x'$ orthogonal to $y'$ and $z'$ is chosen so that its $z$ component equals zero. From figure~\ref{cutplane} it is clear that 
\begin{equation} z = 0 x' + \sin\theta y' + \cos{\theta} z'. \end{equation}
\begin{figure}[h]
\begin{center}
\includegraphics [width=0.3\textwidth]{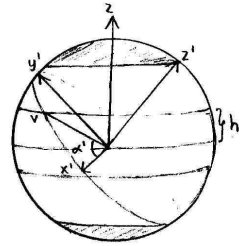} 
\caption{Coordinates $x'$ and $y'$ are introduced in the plane of the great circle orthogonal to the vector $z'$.} \label{sphere}
\end{center}
\end{figure}\\
Meanwhile, as can be seen from figure~\ref{sphere}, a vector $v$ lying just on the boundary of the white belt can be expressed in terms of $y'$ and $x'$ as \begin{equation} v = \cos{\alpha'} x' + \sin {\alpha'} y' \end{equation}
with $\alpha' = \frac{\alpha}{2}$, its $z$ component being equal to zero.
We also know that the $z$ component of our vector $v$ is just $h$, with $h = {\frac{1}{\sqrt{3}}}$ according to our earlier deliberations.
Taken together, this gives \begin{equation}  v \cdot z = h = (\cos{\alpha'}x' + \sin{\alpha'}y')  \cdot z = \sin{\alpha'} y' \cdot z = \sin {\alpha'}\sin\theta  \end{equation}
so that
\begin{equation} \label{alpha} \alpha = 2 \arcsin \frac{h}{\sin\theta} \end{equation} 
and
\begin{equation} \beta = 8 \arcsin \frac{h}{\sin\theta} - 2 \pi. \end{equation}
When $ \theta < \arcsin{\frac{1}{\sqrt{3}}} $ expression~(\ref{alpha}) for $\alpha$ is not defined; for those angles all of the vectors orthogonal to the black section vector defined by the angle $\theta$ lie within the white section.\\
This enables us to express the fraction of the orthogonal great circle corresponding to every choice of vector $z'$ in terms of the angle $\theta$, making possible integration over all values of $\theta$ and thereby the comparison we have in mind.\\
So, the integrals we will want to evaluate are
\begin{equation} I = 2 \pi \int^{\arcsin{\frac{1}{\sqrt{3}}}}_0 \sin{\theta} d\theta + \int^{\frac{\pi}{4}}_{\arcsin{\frac{1}{\sqrt{3}}}}(8 \arcsin \frac{h}{\sin\theta} - 2 \pi)\sin{\theta} d\theta. \end{equation} 
This sum turns out to equal $1.4572$.\\  
This, multiplied by a combinatorial factor of three because what we considered the first vector could as well have been the second or third, should be compared to the value of the integral \begin{equation} 2 \pi \int _0^{\frac{\pi}{2}} \sin\theta d\theta = 2 \pi. \end{equation} The result is that approximately $69 \%$ of all ordered bases in $\mathbb{R}^3$ can be KS-coloured using the given construction.\\ 
The above considerations for the three-dimensional case can with some modification be applied also in four dimensions. Introducing spherical coordinates $\lbrace \phi, \theta_1, \theta_2 \rbrace $ on the three-sphere, we will start out by finding the intersectional area of the orthgonal two-sphere and the white 'belt'. Let $z'$ denote the black vector, let $z$ be a reference coordinate, and let $y'$ be a vector on the orthogonal two-sphere as in figure~\ref{4D}. \\
\\
The white section is the set of vectors \begin{equation} \label{belt3} \lbrace u : \quad \mid u \cdot z \mid \leq A \rbrace, \quad A = \frac{1}{2}. \end{equation}
For any vector $u$ in this set we have that \begin{equation} \label{orthog} u \cdot z' = 0. \end{equation} \\
Now, let us make the ansatz
\begin{equation} y' = az + b z'. \end{equation}
Normalization together with condition (\ref{orthog}) then gives \begin{equation} a = \frac{1}{\sin{\theta_2}}, \quad b= - \frac{\cos{\theta_2}}{\sin{\theta_2}}. \end{equation}
Also, \begin{equation} u \cdot z' = 0 \Rightarrow u \cdot y' = u \cdot (a z + b z') = a u \cdot z.\end{equation}
So, using (\ref{belt3}), the belt on the orthogonal two-sphere will be the set of vectors \begin{equation} \lbrace v : \quad \mid v \mid \leq B \rbrace, \quad B = aA = \frac{1}{2 \sin{\theta_2}}. \end{equation}  
For $0 \leq \theta_2 \leq \arcsin{\frac{1}{2}}$ the orthogonal two-sphere will lie entirely within the white section. \\
This intersection between the orthogonal two-sphere and the white section on the three-sphere can now be treated in analogy with the previous case. Given a black first vector, when placing the second vector in the white section, the segment of the great circle orthogonal to this second vector on which we can choose the third in order for the fourth to lie in the white section is given by \begin{equation} \gamma = 8 \arcsin \frac{B}{\sin{\theta_1}} - 2 \pi. \end{equation}
Also in analogy with the previous case, all of the orthogonal great circle will be white for $\arccos{B} \leq \theta_1 \leq \arcsin{B}$. 
To summarize, we have integration over the angle $\theta_2$ which runs between $0$ and $\frac{\pi}{2}$, covering the black cap, and the possibilities available for choosing the remaining three vectors are governed by a function of $\theta_2$, obtained from an integration over the angle $\theta_1$ between $\arccos{B}$ and $\frac{\pi}{2}$, that is, over the white section of the two-sphere orthogonal to the first vector specified by $\theta_2$, $B$ being a function of $\theta_2$. \\
To make all of this explicit, we have the following integrals
\begin{equation}  I = 2 \pi \int_{\arccos{B}}^{\arcsin{B}}\sin{\theta_1}d\theta_1 + \int^{\frac{\pi}{2}}_{\arcsin{B}}(8 \arcsin{\frac{B}{\sin{\theta_1}}} - 2 \pi)\sin{\theta_1} d\theta_1 \end{equation}
and, finally 
\begin{equation} 4 \pi \int^{\arcsin{\frac{1}{2}}}_0 \sin^2{\theta_2}d\theta_2 + \int^{\frac{\pi}{4}}_{\arcsin{\frac{1}{2}}} I \sin^2{\theta_2} d\theta_2. \end{equation}
The result when comparing this, multiplied by an overall combinatorial factor of four, to the value of the expression
\begin{equation} 4 \pi \int^{\frac{\pi}{2}}_0 \sin^2{\theta}d\theta \end{equation}
is that $34\%$ of the ordered orthogonal triples in $\mathbb{R}^4$ are properly coloured using the chosen method.
\section{Conclusions}
Generalizing a method of colouring proposed by Appleby in three real dimensions, we have found a lower bound on the area of the $n$-sphere that can be KS coloured, but we are still ignorant as to a sharp upper bound.
In three and four dimensions, we have calculated how large a fraction of all bases the coloured area corresponds to, and the behaviour of the percentage (69 \% in three dimensions and 34 \% in four dimensions) suggests that the asymptotic value for a large number of dimensions might well be zero.
These results are restricted to families of pure states that span a real subspace of complex Hilbert space, and clearly it would be interesting to see this restriction lifted. 
It should also be noted that the main advantage of the Appleby colouring is that it is easily generalized to higher dimensions, but that there is no reason to believe that it is particularly effective. It is also not obvious that the same method of colouring would be maximal in different dimensions, and our lower bound for the upper bound might be considerably sharpened by experimenting with other constructions.\\ 

\noindent \textbf{Acknowledgements}\\
Several useful suggestions for this paper have been provided by an anonymous referee. Thanks are also due to Hans Rullg\aa rd for support in numerical computations.


\begin{thebibliography}{99}
 

\bibitem{KS} S. Kochen and E.P. Specker, 'The Problem of Hidden Variables in 
Quantum Mechanics', J. Math. Mech., Vol. 17, 59-87 (1967)

\bibitem{P} I. Pitowsky, 'Quantum Mechanics and Value Definiteness', Philos. Sci., Vol. 52, 154-156 (1985)

\bibitem{M} D.A. Meyer, 'Finite Precision Measurement Nullifies the Kochen-Specker Theorem', Phys. Rev. Letters, Vol. 83, 3751-3754 (1999)

\bibitem{K} A. Kent, 'Noncontextual Hidden Variables and Physical Measurements', Phys. Rev. Letters, Vol. 83, 3755-3757 (1999)

\bibitem{CK} R. Clifton and A. Kent, 'Simulating quantum mechanics by non-contextual hidden variables', Proc. Roy. Soc. A, Vol. 456, 2101-2114 (2000)








\bibitem{Appleby} D.M. Appleby, 'The Bell-Kochen-Specker theorem', 
\texttt{quant-ph/0308114}; Stud. Hist. Philos. Mod. Phys. 36, 1-28 (2005) 

\end{thebibliography}
\end{document}